\documentstyle[12pt]{article}

\input tcilatex

\QQQ{Language}{
American English
}

\begin{document}

\topmargin 0pt \oddsidemargin=-0.4truecm \evensidemargin=-0.4truecm 
\baselineskip=24pt \setcounter{page}{1} 
\begin{titlepage}     
\vspace*{0.4cm}
\begin{center}
{{\LARGE\bf 
Cosmological Relics of the High Energy Cold Dark Matter Particle in the Universe}}
\vspace{0.8cm}

{\large\bf 
Xue-Qian Li$^{1,3}$  and  Zhijian Tao$^{1,2}$}

\vspace*{0.8cm}

1.  China Center of Advanced Science and Technology,  (World Laboratory)

P.O.Box 8730 Beijing 100080, China.

2. Theory Division, Institute of High Energy Physics, Academia Sinica

Beijing 100039, China

3. Department of  Physics, Nankai University, Tianjin, 300071, China.
\vspace{8pt}

\end{center}
\vspace{2truecm}     

\begin{abstract}
We demonstrate that if the universe is
dominated by the massive cold dark matter, then 
besides the generally
believed thermal distribution of the dark matter relics, there may exist
some 
very energetic non-thermal relics of the dark matter particles in the
universe from some unknown 
sources, such as from decay of supermassive X particle released from 
topological defect collapse or annihilation. 
Very interesting, we point out that these high energy 
dark matter particles
may be observable in the current and future cosmic ray experiments. 
\end{abstract}
\end{titlepage}
\renewcommand{\thefootnote}{\arabic{footnote}} \setcounter{footnote}{0}

There is a great deal of evidence to indicate that the dominant component of
the universe matter density is dark matter. The most direct evidence for
existence of dark matter is from the astronomical observation of velocities
of the spiral galaxies \cite{kol}. In our own galaxy, for example, in order
to explain the observation, a local dark matter density $\rho ^{DM}=0.3{\rm %
GeV}/{\rm cm}^3$ is determined \cite{data}. In addition, the inflation
theory predicts a flat universe, i.e. $\Omega =1$. The standard big-bang
nucleosynthesis implies that the ordinary matter can contribute at most 15
percent of the critical density. It means that 90 percent of the matter in
our universe may be dark \cite{tur}. The most attractive candidate of the
dark matter is from the particle physics. One of them is the
weakly-interacting massive particle (WIMP). The WIMP serves as a natural
cold dark matter candidate to the universe, since if the WIMP is at a
weak-scale mass it naturally provides a near closure density to the present
universe as required by the inflation theory. In the present universe the
WIMP dark matter particle is statically distributed. Its density is
controlled by the Boltzman equation and the evolution of the universe, so it
is called thermal relic abundance. So far experimental effort to search for
the cold dark matter is made to find the signals of this distribution. The
most promising experiments involve the direct detection at low background
detectors, which look for the signal of dark matter particle as it collides
with detector matter, and the indirect detection through observation of
energetic neutrinos emerging from annihilation of cold dark matter particles
which are accumulated in the Sun and the Earth \cite{kol}. Having not seen
any signal in these experiments the parameter space of the WIMP models is
seriously restricted. However to discover the WIMP dark matter particles or
rule out the WIMP models much more efforts are needed, hopefully the next
generation detectors will cover much larger parameter space of the WIMP
models.

Instead, in this letter we discuss another possibility to search for cold
dark matter. We start from a question, whether there exists a significant
flux of very energetic cold dark matter particle in the universe. Since the
thermal relics of cold dark matter are statically distributed in the
universe, we are actually studying some non-thermal relics. It is believed
that if heavy particles are produced at a temperature of the universe which
is not much lower than the mass scale of the particle, the number of the
high energy relics would reduce very fast due to thermalization processes
and the expansion of the universe. As a result at present time there would
be no heavy particles with high energies. However, there indeed is a
possibility of existence of non-thermal relics at present universe. Its
reason is not well understood, but they may wander around at present time.
For example, at a later stage of the universe evolution when its temperature
is sufficiently low, some topological defects, such as monopoles, cosmic
strings are decoupled from the thermal bath of the universe. Later the
superheavy fields, which form the defects, decay into some heavy particles.
Since the parents are out of thermal equilibrium, the products are also
non-thermally distributed over the space. That could serve as a new source
for such non-thermal relics in our universe. Quite similar to the fact that
there exists large flux of high energy cosmic ray components of nucleon and
other stable particles of the standard model, it is natural to expect an
existence of energetic cold dark matter particle in the cosmic rays as the
dark matter particles dominate the universe density. The question is how
much this kind of energetic dark matter particles remain in our universe. If
the amount of the particles are very small, it is not interesting at all to
study the potential of search for them. By a general and reasonable analysis
we are suggesting there may exist a large amount of this kind of particles
in the universe. Since the WIMP is electrically neutral and heavy, so the
ordinary accelerating mechanisms for nucleon and neutrino in the cosmic rays
cannot work for WIMP's. Two mechanisms for generating energetic cold dark
matter particles in the universe are considered below. One is through the
collision of cosmic ray particles and statically distributed dark matter
particles. Since the energy of the cosmic ray ranges from low to very high,
the collision will result in some high energy cold dark matter particle
fluxes. Another one is that high energy cold dark matter particles can also
originate from the very early universe. Some examples are the decays of
superheavy X particle released from destruction of topological defects
formed during phase transition in the very high energy scale, like the grand
unification scale. Recently some decays of the superheavy X particles with
mass close to the grand unification scale released from topological defect
to standard model light particles have been interpreted as a new mechanism
to explain the observed extremely high energy cosmic rays, though it seems
that the simple version of the topological defect model is not able to fit
all the observed cosmic ray spectrum \cite{bla,bha}. Similarly, the very
massive X particle may also decay to cold dark matter particles resulting in
the energetic particles in the universe. By considering these two sources we
derive an evolution equation for the flux of the energetic cold dark matter
in the universe. We assume an equilibrium solution to the equation, then we
find that the flux of the high energy WIMP's is sizable, therefore it should
be detectable in some future large detectors. We stress that our scenario
can also provide a possible explanation for the exotic event observed in
Yunnan Cosmic Ray Station (YCRS) \cite{yun,che}, if it is confirmed.

It is a very interesting question to derive the effective density or flux of
the high energy neutral WIMP particle in the present universe. We denote $E$
as the kinetic energy of the WIMP. For the static distribution of the
thermal relic WIMP's, $E=0$. Our task is to derive the density of WIMP's
with $E\not =0$, say from GeV to much higher energy. We assume that there
already exists a non thermal relic density of WIMP's with nonzero kinetic
energy $n(E)$. As we stated, this density may result from some early
universe sources, like decays of X particle released from topological
defect. In principle , the scattering between the WIMP's (including both
static thermal and high energy non-thermal relics) and the energetic cosmic
protons can make a new equilibrium with the existing high energy density $%
n(E)$.

We denote the differential WIMP density in the energy region from $E_2$ to $%
E_2+\Delta E_2$ as $\Delta n(E_2)$. The scattering cross section $%
\sigma(E_1,E_2,E_1^{\prime},E_2^{\prime})$ for a cosmic proton and WIMP into
a WIMP plus other final state particles can be estimated, provided the model
of WIMP interaction is given, here we just keep in mind this scattering is a
weak interaction process. The equation for the distribution evolution of $%
n(E_2)$ is derived as the following

\begin{eqnarray}
{d\Delta n(E_2)\over dt} &=& -\int {d\sigma (E_1,E_2,E_1',E_2')\over dE_2'}
dE_2'
{dF(E_1)\over dE_1}dE_1\Delta n(E_2)(1-\delta_{E_2E_2'}) \nonumber\\
&+& \int 
{d\Delta\sigma (E_1,E_2'',E_1',E_2)\over dE_2''}dE_2''
{dF(E_1)\over dE_1}dE_1n(E_2'')
(1-\delta_{E_2''E_2})
\end{eqnarray}
where $E_1^{\prime },~E_2^{\prime }$ are the kinetic energies of the
outgoing hadron and WIMP, $dF(E_1)/dE_1$ is the energy density of the
charged cosmic ray particle flux and $\Delta \sigma (E_1,E_2^{\prime \prime
},E_1^{\prime },E_2)={\frac{d\sigma (E_1,E_2^{\prime \prime },E_1^{\prime
},E_2)}{dE_2}}\Delta E_2$. The first term on the left-hand side of the
equation reduces $\Delta n(E_2)$ because a certain number of dark matter
particles with energy from $E_2$ to $E_2+\Delta E_2$ are stricken out of the
energy range by cosmic ray collision. On the other hand some dark matter
particles with energy outside the range are stricken into the range by the
cosmic ray collision. So the second term increases $\Delta n(E_2)$. We
notice that the incident cosmic ray flux should include all possible sources
in cosmic rays, charged particles, cosmic photons, neutrino background and
etc. However the cross section of the WIMP and the very low temperature
photon or neutrino scattering is much suppressed, so the effect can be
neglected here. For the same reason we also neglect the contribution of
cosmic rays with energy much lower than 1 GeV. The flux of energetic charged
cosmic ray particles is measured on the earth and it has the spectrum in the
outer space as $F(\Lambda )(E_1/\Lambda )^{-2.7}$ while $0.4{\rm GeV}\leq
E_1\leq 10^6$ GeV, the parameter $\Lambda $, $F(\Lambda )$ and the spectrum
form for higher energy can be found in reference \cite{gai}. For our purpose
the most relevant energy range should not be much higher than the weak
scale, since for very high energy scale both the cosmic ray flux and the
initial density $n(E_2)$ are too much suppressed (see below).

Obviously, after infinitely long evolution time, $d\Delta n(E_2)/dt=0$ to
reach an equilibrium. However for a finite life time of the universe as long
as $10^{10}$ years, the equilibrium is not necessarily reached. Our
investigation is divided into two steps. Here we first assume the
equilibrium condition is satisfied. We will come back to this point later.
Moreover since the dominant density of dark matter is the static thermal
relics, as a very good approximation one may use $n(E_2^{\prime\prime})=n(0)%
\delta_{E_2^{\prime\prime}0}$ in the above equation so one has 
\begin{equation}
\frac{\Delta n(E_2)}{\Delta E_2}= \frac{n(0)\int {\frac{d\sigma(E_1,0,E_1^{%
\prime},E_2) }{dE_2}} {\frac{dF(E_1)}{dE_1}}dE_1} {\int \sigma(E_1,E_2){\ 
\frac{dF(E_1)}{dE_1}}dE_1}, 
\end{equation}
where $n(0)$ is the local Galactic halo density contributed by the WIMP dark
matter, and $\sigma(E_1,E_2)$ means that all possible final state energies
have been integrated over.

So long as the equilibrium condition is satisfied, the equilibrium solution
indicates that the differential spectrum of WIMP's in our galaxy is
proportional to the local halo density. And it is independent of the details
of the mechanism of the early universe evolution. No matter what the
mechanism it is, it only needs to provide certain large initial $n(E_2)$ for
the evolution equation, so that the equilibrium condition can be satisfied.

Without using detail information of any WIMP models we can very roughly
estimate the equilibrium differential density spectrum of WIMP's with large
kinetic energy. If $E_2$ is much smaller than $10^6$ GeV while larger than a
certain value, say 1 GeV, simply by dimension analysis one estimates 
\begin{equation}
\frac{dn(E_2)}{dE_2}\sim \frac{n(0)}{E_2}(\frac{1GeV}{E_2})^{2.7} 
\end{equation}
this corresponds to a flux $j(E_2\ge 10^2GeV)\sim 1{\rm cm}^{-2}{\rm s}^{-1}$%
. Though this flux is much smaller than the static dark matter flux across
the Earth, which is about $10^5{\rm cm}^{-2}{\rm s}^{-1}$. This high energy
WIMP flux may still cause some detectable signals in future large detectors,
since it is a high energy involved process. To calculate the capture rate in
the high energy process, one needs to specify a concrete WIMP model. However
since the capture process is typically a weak interaction process, by
assuming a weak interaction cross section as $10^{-35}-10^{-38}{\rm cm}^2$,
we can reasonably estimate a capture rate as $10^{-5}-10^{-8}{\rm s}^{-1}$
per ton of target. The expected events in one year are 100 and 0.1
respectively.

There could be many new signatures to distinguish the desired processes from
other background processes. We mention one possible signature which is to
produce a heavy charged particle when the neutral dark matter hitting on the
matter surrounding or inside the detector. For instance in SUSY model, if
neutralino is the dark matter particle, there exist some charged
supersymmetric particles. The mass difference of the charged particle and
the neutralino should be comparable with the mass of neutralino itself, so
if the incident kinetic energy of the neutralino is larger than the mass
difference, the charged supersymmetric particle could be produced. Because
the charged particle is heavy and unstable, it decays to some light products
afterwards. The kinematics and decay properties of the charged particles may
provide some clues for the incident dark matter particles.

In 1972, in the cloudy chamber of the YCRS, an exotic event with a heavy
charged particle being tracked was observed \cite{yun}. In fact, three
charged prongs were recorded and they correspondingly possessed three$-$%
momenta as $p_a=6.6_{-0.8}^{+10}$GeV/c, $p_b=62$GeV/c and $p_c=110$GeV/c
with the spanned angles as $\theta _{ab}=3^{\circ }25\prime $, $\theta
_{bc}=1^{\circ }25\prime $and $\theta _{ac}=4^{\circ }55\prime $, namely
they were coplanar trajectories. Later the tracks $a$ and $b$ were
identified as $\pi ^{-}$ and proton while $c$ remained unknown.. A careful
re-analysis which was done recently by Chen {\it et al.}\cite{che} indicates
that the unidentified positively charged particle has a heavy mass as 
$$
M^{+}>43\;{\rm GeV} 
$$
The authors of reference \cite{che} have suggested that the unknown charged
heavy particle may be identified as a heavy elementary particle which was
produced by a bombard of a heavy neutral particle coming from the universe
on a proton. Obviously, in this case, the kinetic energy of the unknown
neutral particles must be greater than the threshold $M^{+}-M^0+M_\pi$ where 
$M^0$ is the mass of the neutral particle. For so large momenta measured in
the case, the neutral particle must be very energetic. If their postulation
is valid, it would be a direct evidence for the existence of large flux of
massive and high energy neutral particle in the cosmic ray, even though with
a remarkable uncertainty. They also claimed to have found similar evidence
at the data obtained by other experiments \cite{che}. Numerically the flux
should be around $10 {\rm cm}^{-2}{\rm s}^{-1}$ in order to explain the
observed capture rate. The question was where the high energy neutral
particle flux is from. We see that the high energy WIMP flux derived above
can give the necessary amount of flux.

Now we come back to discuss the equilibrium condition which is assumed
before. We notice that an initial condition must be satisfied, otherwise the
equilibrium cannot be reached for the present age of our universe. The
condition is that the initial density $n(E_2)$ with $E_2$ being greater than
a certain value is large enough. If one assumes that the initial $n(E_2)$ is
negligibly small, from the evolution equation we can roughly estimate the
order of the magnitude of the density, which is many order of the magnitudes
smaller than the equilibrium density we obtained above. It means that by the
collision of the high energy cosmic ray and the static WIMP dark matter
particle alone, the produced high energy WIMP density in the universe is
negligible. Therefore we need a large initial condition for the evolution
equation. This implies that some new sources for generation of such initial
condition should exist. We have named some possibilities such as decay of X
particle released from topological defect\cite{bla}. Here the question is if
these new mechanisms are indeed able to produce large enough $n(E_2)$ in the
early universe. Certainly we are not trying to investigate all the
possibilities, but considering only the case of cosmic string collapse \cite
{bha1} and annihilation of monopole-antimonopole pairs \cite{bha2}.
Following reference \cite{bha}, the rate of release of massive particle X
forming the topological defect from destruction of the defect can be
expressed as 
\begin{equation}
\frac{dn_X(t)}{dt}=km_Xt^{-3} 
\end{equation}
where $m_X$ is the mass of the X particle, $t$ is the Hubble time, and $k$
is a dimensionless constant which is restricted to be less than about 0.2 by
the cosmic ray spectrum observation \cite{bha}. It corresponds to $p=1$ in
the equation (1) of reference \cite{bha}. Moreover, by assuming the massive
particle X decay with a sizable branching ratio to stable WIMP particles, we
can use the same formula of equation (5) in this reference for the
differential particle flux $j(E_0)$. For the weak interaction feature of the
WIMP we omit the energy loss effects when the particle propagates in the
universe from some early time $t_i$ to the present time $t_0$, namely $%
dE_i/dE_0$ is dominantly determined by the expansion effect of the universe
in the equation, here $E_i$ and $E_0$ are the energies of the particle at
time $t_i$ and $t_0$ respectively. Then the flux is estimated as 
\begin{equation}
E_0\frac{dj(E_0)}{dE_0}\simeq ckm_X(\frac{m_X}{E_0})t_0^{-2} 
\end{equation}
where $c$ is the speed of light and $t_0$ is the age of the universe. Taking 
$m_X=10^{16}$ GeV, and $k=0.1$ we obtain the differential flux due to the X
particle decay as 
\begin{equation}
E_0\frac{dj(E_0)}{dE_0}\sim \frac{100{\rm GeV}}{E_0}{\rm cm}^{-2}{\rm s}%
^{-1} 
\end{equation}
This flux is comparable with the equilibrium flux obtained above, so we
conclude that it is possible to satisfy the equilibrium condition with the X
particle decay as the source of high energy WIMP density in the early
universe.

In this work we suggest to directly detect WIMP dark matter particles in the
cosmic ray. The point is that we show that there may exist some sizable flux
of high energy WIMP particles in the universe. Even though the flux is much
smaller than that occurring in the direct dark matter search experiments,
the detector for high energy experiments can be much bigger than that built
for the low energy direct search experiments, therefore the numbers of
events in high energy experiments are not necessarily less than that in the
low energy experiments. And the YCRS event may be an indication of this kind
of signal, even if the measurement on the event is not very reliable and
needs new experiments to check. In fact some cosmic ray experiments in L3
are undergoing, one of the experimental purposes is to search for this kind
of events \cite{ho}. Once the characteristics of those events are confirmed,
the observation can be viewed as a strong evidence supporting the existence
of the stable WIMP's of high energy in the universe.

In order to understand where the high energy WIMP dark matter particles come
from, we have derived an evolution equation for the differential density.
With assumption of the equilibrium condition we estimate the differential
density numerically and find that this density is interestingly large for
future and even some current cosmic ray experiments. The interesting
property of the equilibrium is that it does not depend on the details of any
sources in the early universe, so long as the equilibrium condition is
satisfied. We use the grand unification scale supermassive X particle
released from $p=1$ type topological defect decay as an example to show that
the equilibrium is possibly reached. One should have noticed that all the
numerical estimation in this work is somehow crude, they may be one or two
orders of magnitudes deviating from the stated values. And the decay mode of
X particle is not clear, while we simply assume that it decays dominantly to
the WIMP dark matter particle. Though there are those uncertainties,
however, we believe that the qualitative features we discuss in this work
are correct and may be very interesting to experimentalists. Finally we
emphasize that the possibility of existence of high energy cold dark matter
particle in the universes deserves much attention of both theoreticians and
experimentalists. If the equilibrium state is not reached in the real world,
so the flux is not as large as the equilibrium result we derived , the
desired flux or density of the cold dark matter particle in the universe
from some new sources, like X particle decay from topological defect
destruction, can still be large enough for the future experimental detection.

It is a pleasure to express our gratitude to C. Qing and T. Ho for useful
discussion. This work is partly supported by the National Natural Science
Foundation of China (NSFC).

\end{document}